# Improved temperature dependence of rate coefficients for rotational state-to-state transitions in $H_2O + H_2O$ collisions

Bikramaditya Mandal and Dmitri Babikov[*]

*Chemistry Department, Wehr Chemistry Building, Marquette University,*
*Milwaukee, Wisconsin 53201-1881, USA*

**ABSTRACT**

*Aim:* We present an improved database of temperature dependent rate coefficients for rotational state-to-state transitions in $H_2O + H_2O$ collisions. The database includes 231 transitions between the lower para- and 210 transitions between the lower ortho-states of $H_2O$ (up to $j = 7$) and can be employed for cometary and planetary applications up to the temperature of 1000 K.

*Methods:* New general method is developed and applied which permits to generate rate coefficients for excitation and quenching processes that automatically satisfy the principle of microscopic reversibility and, also, helps to cover the range of low collision energies by interpolation of cross sections between the process threshold and the computed data points.

*Results:* It is found that in the range of intermediate temperatures, $150 < T < 600$ K, new rate coefficients are in good agreement with those reported earlier, but for higher temperatures, $600 < T < 1000$ K, the new revised temperature dependence is recommended. The low temperature range, $5 < T < 150$ K, is now covered, by the abovementioned interpolation of cross sections down to the process threshold.

[*] Author to whom all correspondence should be addressed; electronic mail: dmitri.babikov@mu.edu

## 1. Introduction

Recently, a database intended for the modelling of cometary comas and planetary atmospheres, was reported, that includes thermally averaged cross sections (TACS) for rotational state-to-state transitions in $H_2O + H_2O$ collisions (Mandal and Babikov, 2023). That database contains 231 transitions between para-states and 210 transitions between ortho-states of a "target" water molecule (for $H_2O$ states at energies below $E = 700$ cm$^{-1}$, up to $j = 7$), obtained assuming thermal distribution of the rotational states of "collider" water molecules, in the temperature range between 150 K and 800 K. It was also assumed that cross sections $\sigma$ from that database can be easily converted into the corresponding rate coefficients using $k(T) = v_{ave}(T)\,\sigma(U)$, where $v_{ave}(T) = \sqrt{8k_BT/(\mu\pi)}$ is average collision velocity at given temperature, $\mu$ is the reduced mass of collision partners, and $U = (4/\pi)\,k_BT$ is kinetic energy which corresponds to the average collision velocity.

It should be understood, though, that this simple method of computing rate coefficients is approximate, because it assumes that one value of collision energy $U$ can represent thermal distribution, which is equivalent to assuming that cross section $\sigma$ is independent of $U$. Although it was used in the recent literature (Boursier et al., 2020; Buffa et al., 2000), this method is expected to be valid only for a short range of intermediate temperatures. Indeed, it may break down at lower temperatures because at low collision energies the energy dependence of cross section $\sigma(U)$ is usually sharper, and/or at higher temperatures when the energy distribution is broader, with no one representative value of $U$.

The rigorous expression for calculation of rate coefficient includes averaging over thermal distribution of collision energies (for $n \rightarrow n'$ quenching process):

$$k_{n \rightarrow n'}(T) = \frac{v_{ave}(T)}{(k_BT)^2} \int_{E=0}^{\infty} E\,\sigma_{n \rightarrow n'}(E)\,e^{-\frac{E}{k_BT}}dE, \qquad (1)$$

where $\sigma_{n \rightarrow n'}(E)$ is the actual dependence of inelastic cross section on the kinetic energy of collision (Żółtowski et al., 2022). The first, applied goal of this paper, is to employ formulae like Eq. (1), which improves the overall prediction of temperature dependence of rate coefficients but also allows permits expanding the range of temperatures towards the low-$T$ regime. This is particularly important for the modelling of cometary environments (Bockelée-Morvan et al., 2010;

Cochran et al., 2015; Dones et al., 2015) (Cordiner et al., 2020) or the atmospheres of icy planets (Hartogh et al., 2011; Vorburger et al., 2022; Wirström et al., 2020), where the temperatures below 150 K are typical. The expansion of $H_2O + H_2O$ database into the low-$T$ range is achieved here by interpolation between the reference data (Mandal and Babikov, 2023) and the process threshold, where the kinetic energy is barely enough for the excitation process $n' \rightarrow n$ and thus the excitation cross section $\sigma_{n' \rightarrow n}(E)$ is known to vanish. Note that atmospheres of Ganymede, Calisto and Europa will be investigated in detail by ESA's Jupiter Icy Moons Explorer mission (JUICE) focused on rotational transitions in water at high resolution (Bruzzone et al., 2013; Enya et al., 2022; Grasset et al., 2013; Vorburger et al., 2022). Moreover, in cometary comas non-equilibrium conditions occur between the inner (collision-controlled) and the outermost (fluorescence-controlled) parts of the expanding gas (Bonev et al., 2021; Cochran et al., 2015; Cordiner et al., 2022; Roth et al., 2021). Interpretation of these and many other observations require numerical modeling with radiation transfer codes which, in turn, requires collision rate coefficients for $H_2O + H_2O$ as input. Large errors of these rate coefficients can affect the predictions of astrophysical modeling by orders of magnitude (Al-Edhari et al., 2017; Faure et al., 2018; Faure and Josselin, 2008; ben Khalifa et al., 2020).

The second, more methodological goal of this paper, is to offer a general method for the calculations of rate coefficients that satisfy the principle of microscopic reversibility:

$$w_n \, k_{n \rightarrow n'} = w_{n'} \, k_{n' \rightarrow n} \qquad (2)$$

Here $w_n$ and $w_{n'}$ are statistical populations of the two states:

$$w_n = (2j+1) e^{-E_n / k_B T} / Q,$$

$$w_{n'} = (2j'+1) \, e^{-E_{n'} / k_B T} / Q,$$

$E_n$ and $E_{n'}$ are their energies, $2j+1$ and $2j'+1$ are corresponding rotational degeneracies and $Q = \sum w_n$ is rotational partition function. What is the meaning of Eq. (2)? It simply tells us that at equilibrium the transfer of populations in two directions, $n \rightarrow n'$ and $n' \rightarrow n$, are equal, which means that if the system reached equilibrium it will remain in the dynamic equilibrium. Even if the system is not in the equilibrium, the rate coefficients $k_{n \rightarrow n'}$ and $k_{n' \rightarrow n}$ must satisfy Eq. (2) because, if they don't, the system will never come to the equilibrium, which is unphysical.

Thus, Eq. (2) is a fundamental property, required in any modeling of state-to-state transition kinetics. It is automatically fulfilled by rigorous quantum mechanical calculations of rate coefficients (Daniel et al., 2011; Dubernet and Quintas-Sánchez, 2019; Faure et al., 2020; Sun et al., 2020; Wiesenfeld et al., 2011; Wiesenfeld and Faure, 2010), but is not immediately satisfied by approximate methods, such as a semiclassical method (Boursier et al., 2020; Buffa et al., 2000) or a mixed quantum/classical theory, MQCT (Mandal and Babikov, 2023). Unfortunately, rigorous full-quantum calculations are affordable only for smaller (lighter) molecules and lower collision energies (temperatures). In particular, they are not possible for $H_2O + H_2O$ collisions in the temperature range of interest. Therefore, one must rely on the approximate methods referenced above, that require some symmetrization (Boursier et al., 2020) of computed rate coefficients for the excitation and quenching directions of each transition $n' \leftrightarrow n$, before those could be used in the modelling of kinetics. Here we devise a new (to our best knowledge) method of symmetrization and employ it to produce a new database of $H_2O + H_2O$ rate coefficients using cross sections from the MQCT database reported previously (Mandal and Babikov, 2023). In principle, this symmetrization approach can be used in conjunction with any other input data, including the results of classical trajectory simulations (Faure et al., 2005; Loreau et al., 2018; Wiesenfeld, 2021).

## 2. Details of the method

The principle of microscopic reversibility is usually expressed in terms of thermal rate coefficients for excitation and quenching directions of a transition, as given by Eq. (2). However, it is also important to derive two other versions of the principle: in terms of cross sections, and, in terms of transition probabilities of individual trajectories. For that, we should first write down the analogue of Eq. (1), but for the excitation process:

$$k_{n'\to n}(T) = \frac{v_{ave}(T)}{(k_B T)^2} \int_{E'=\Delta E}^{\infty} E' \, \sigma_{n'\to n}(E') \, e^{-\frac{E'}{k_B T}} dE', \qquad (3)$$

Here $E'$ denotes the kinetic energy of collision in the case of excitation. The integration starts from the excitation threshold $\Delta E = E_n - E_{n'}$ because at energies below $E' < \Delta E$ the value of excitation cross section $\sigma_{n'\to n}$ is zero. Substitution of Eq. (1) and (3) into Eq. (2), and the change of integration variable from $E'$ to $E = E' - \Delta E$, leads to the following result:

$$\frac{E \, \sigma_{n\to n'}(E)}{2j'+1} = \frac{E' \sigma_{n'\to n}(E')}{2j+1} \qquad (4)$$

This is the principle of microscopic reversibility in terms of cross sections. Note that in this equation the excitation and quenching cross sections are computed at different collision energies, but the same total energy (kinetic energy of collision plus internal rotational energy):

$$E_{tot} = E_n + E = E_{n'} + E'.$$

This expression is consistent with $E = E' - \Delta E$ used to obtain Eq. (4), and with different ranges of integration in Eqs. (1) and (3). Namely, when the value of $E'$ is at its lowest limit $E' = \Delta E$, the value of $E$ is also at its lowest limit, $E = 0$.

Again, Eq. (4) is usually satisfied automatically by the rigorous quantum calculations of cross sections. What is the meaning of Eq. (4)? Recall that cross sections $\sigma$ are, normally, summed over the final degenerate states (projections of angular momentum $j$ onto the axis of quantization). There are $2j' + 1$ of these final states for transition $n \to n'$ and $2j + 1$ final states for transition $n' \to n$, which means that the excitation and quenching cross sections in Eq. (4) are divided by the number of final states. This makes sense because the values of $j$ and $j'$ can be very different, and the degeneracies of the two states can differ significantly. Equation (4) requires only that the average values of these cross sections (per one final state) are equal.

Now recall that our goal is to use an approximate MQCT method, where cross sections are computed using the following semiclassical equation (in fact, a very similar equation is used in the full quantum calculations):

$$\sigma_{n \to n'} = \frac{\pi}{k^2} \frac{1}{2j+1} \sum_{\ell=0}^{\ell_{max}} (2\ell+1) \, p_{n \to n'}^{(\ell)} \tag{5}$$

where summation is over MQCT trajectories labeled by $\ell$ (the initial orbital angular momentum of collision partners), $p_{n \to n'}^{(\ell)}$ represents the probability of $n \to n'$ transition computed for each trajectory, $k$ is wave vector of collision, $2j + 1$ takes care of averaging over the initial degenerate states. Equation (5) can be substituted into the left- and right-hand sides of Eq. (4). Since $E = (k\hbar)^2/(2\mu)$ and $E' = (k'\hbar)^2/(2\mu)$, where $\mu$ is the reduced mass of collision partners, most factors cancel, leading us to the final important result:

$$p_{n \to n'}^{(\ell)}(E) = p_{n' \to n}^{(\ell)}(E') \tag{6}$$

This is the principle of microscopic reversibility in terms of transition probabilities, which tells us that those must be equal for the individual trajectories describing the processes of excitation and quenching.

It was recognized long time ago (Billing, 1984) that, if we launch two different trajectories, one with the initial state $n$ and collision energy $E$, and another with the initial state $n'$ and collision energy $E'$, we can hope to satisfy Eq. (6) only in the classical limit $\Delta E = 0$, when the values of collision energies $E$ and $E'$ become equal. This is so because both excitation and quenching processes are driven by the same off-diagonal matrix element for $n \leftrightarrow n'$ transition, but the value of transition probability depends on collision velocity too (Billing, 2003). If $\Delta E \neq 0$, the two trajectories will have different collision energies $E$ and $E'$, will proceed with two different collision velocities:

$$v = \sqrt{2\mu E} \text{ and } v' = \sqrt{2\mu E'}, \tag{7}$$

and will result in two different transition probabilities, violating Eq. (6). This is particularly so for the vibrational transitions with large $\Delta E$, but is also true in general, including the rotational state-to-state transitions.

Gert Billing offered a simple solution to this issue (Billing, 2003). He defined the *effective* collision energy $U$ using an average of collision velocities for excitation and quenching processes:

$$\frac{v + v'}{2} = \sqrt{2\mu U}, \tag{8}$$

and proposed to run both excitation and quenching trajectories with this one value of collision energy. The resultant transition probabilities approximately satisfy the principle of microscopic reversibility,

$$p_{n \to n'}^{(\ell)}(U) \approx p_{n' \to n}^{(\ell)}(U)$$

because the effect of the initial state ($n$ vs $n'$) on the trajectory of scattering is relatively weak. Moreover, cross sections obtained from these probabilities approximately satisfy Eq. (4) if one relates the values of $E$ and $E'$ to $U$ using its definition. Namely, substituting Eqs. (7) into (8) and solving for $E$, or alternatively for $E'$, gives:

$$E = U + \frac{\Delta E}{2} + \frac{(\Delta E)^2}{16U}, \tag{9}$$

$$E' = U - \frac{\Delta E}{2} + \frac{(\Delta E)^2}{16U}. \tag{10}$$

We see that these values of $E$ and $E'$ are shifted to the right and to the left from $U$ respectively: $E' < U < E$. In the limit of high collision energy, $U \gg \Delta E$, the value of $U$ is in the middle between $E$ and $E'$. In the classical $\Delta E \to 0$ limit we have simply $U = E = E'$. The condition $E = E' - \Delta E$ is always satisfied. One can also conclude that when the values of $E$ and $E'$ approach their lower limits ($E \to 0$ and $E' \to \Delta E$) the value of $U$ also approaches its threshold, $U \to \Delta E/4$.

This means that in the reversible semi-classical calculations the value of effective collision energy $U$ should never be less than a *quarter* of the excitation quantum $\Delta E$, but the behavior near threshold of both excitation process (near $E' = \Delta E$) and quenching process (near $E = 0$) can be obtained from such calculations. This conclusion is not at all obvious and may even appear counterintuitive at first look. However, the method of Billing permits us to obtain cross sections $\sigma_{n \to n'}(E)$ and $\sigma_{n' \to n}(E')$ that satisfy approximately the principle of microscopic reversibility, exhibit a correct behavior near threshold and agree reasonably well with the results of full quantum calculations. Indeed, the property of reversibility was tested in detail for the ro-vibrational quenching of CO(v=1) + He (Semenov et al., 2013), rotational excitation of $C_6H_6$ by He impact (Mandal et al., 2022), and rotational energy exchange in $ND_3$ + $D_2$ collisions (Joy et al., 2023). Moreover, very good agreement with full quantum results was reported for $H_2$ + He (Semenov and Babikov, 2014), $N_2$ + Na (Semenov and Babikov, 2013), $H_2$ + $H_2$ (Semenov and Babikov, 2016), $N_2$ + $H_2$ (Semenov and Babikov, 2015a), $H_2O$ + He (Semenov et al., 2014), $H_2O$ + $H_2$ (Mandal et al., 2020) and $HCOOCH_3$ + He (Semenov and Babikov, 2015b).

Using this method of computing cross sections, one can also compute rate coefficients $k_{n \to n'}$ and $k_{n' \to n}$ for quenching and excitation processes that satisfy the principle of microscopic reversibility exactly, but his can be done in several different ways. For the sake of completeness, we outline all of them below.

## 2a. Rate coefficients using quenching (Method 1)

Here the idea is to keep only one integration variable $E$ (that corresponds to quenching) in both Eqs. (1) and (3). Equation (1) is already in an appropriate shape. Equation (3) can be rewritten by changing the variable from $E' = E + \Delta E$ to $E$:

$$k_{n'\to n}(T) = \frac{v_{ave}(T)}{(k_BT)^2} e^{-\frac{\Delta E}{k_BT}} \times \int_{E=0}^{\infty} (E+\Delta E)\sigma_{n'\to n}(E+\Delta E) e^{-\frac{E}{k_BT}} dE.$$

Note that the range of integration is now the same as in Eq. (1), the Boltzman exponent is also the same, but the expression got a pre-factor $e^{-\frac{\Delta E}{k_BT}}$. Now let's introduce the following function:

$$F(E,T) = (2j+1)E\, \sigma_{n\to n'}(E)\, e^{-\frac{E}{k_BT}} \tag{11}$$

$$= (2j'+1)(E+\Delta E)\, \sigma_{n'\to n}(E+\Delta E)\, e^{-\frac{E}{k_BT}}.$$

The first and second lines of this equation are equal, due to the principle of microscopic reversibility, Eq. (4). Using this definition, the rate coefficients are computed as:

$$k_{n\to n'}(T) = \frac{v_{ave}(T)}{(k_BT)^2} \left(\frac{1}{2j+1}\right) \int_{E=0}^{\infty} F(E,T)\, dE, \tag{12}$$

$$k_{n'\to n}(T) = \frac{v_{ave}(T)}{(k_BT)^2} \left(\frac{e^{-\frac{\Delta E}{k_BT}}}{2j'+1}\right) \int_{E=0}^{\infty} F(E,T)\, dE. \tag{13}$$

Since Eqs. (12) and (13) are expressed through the same integral, the two rate coefficients satisfy the principle of microscopic reversibility exactly, see Eq. (2). The approximate nature of the semiclassical method is now hidden in Eq. (11), where the first and second lines may not be equal exactly, but only approximately equal.

It becomes clear that the best practical approach is to use both quenching and excitation cross sections $\sigma_{n\to n'}(E)$ and $\sigma_{n'\to n}(E+\Delta E)$ to build one integrand function $F(E,T)$, as a compromise, even if they don't satisfy Eq. (4) exactly. But then, Eq. (2) is automatically satisfied using Eqs. (12) and (13).

## 2b. Rate coefficients using excitation (Method 2)

One can do the same trick using $E'$ as a variable (that corresponds to excitation) and changing the integration limits in Eq. (1), which leads to the following definition of the integrand function:

$$F(E',T) = (2j+1)(E'-\Delta E)\sigma_{n\to n'}(E'-\Delta E)\, e^{-\frac{E'}{k_BT}} \tag{14}$$

$$= (2j'+1)\, E'\, \sigma_{n'\to n}(E')\, e^{-\frac{E'}{k_BT}},$$

and the following expressions for rate coefficients

$$k_{n \to n'}(T) = \frac{v_{ave}(T)}{(k_B T)^2} \left( \frac{e^{+\frac{\Delta E}{k_B T}}}{2j+1} \right) \int_{E'=\Delta E}^{\infty} F(E', T) dE', \quad (15)$$

$$k_{n' \to n}(T) = \frac{v_{ave}(T)}{(k_B T)^2} \left( \frac{1}{2j'+1} \right) \int_{E'=\Delta E}^{\infty} F(E', T) dE'. \quad (16)$$

Comparing these formulae to Eqs. (12-13) one notices that, besides a different integration limit, a different pre-factor $e^{+\Delta E/k_B T}$ (with positive sign) show up in a different spot. Still, if one integrand function $F(E', T)$ is built using the information from both quenching and excitation cross sections $\sigma_{n \to n'}(E' - \Delta E)$ and $\sigma_{n' \to n}(E')$ computed approximately, then Eqs. (15) and (16) will satisfy Eq. (2) exactly.

## 2c. A symmetrized approach (Method 3)

Methods 1 and 2 described above give the main idea, but our actual goal is to use as an integration variable the effective energy $U$ introduced by Billing, rather than either $E$ or $E'$. This is done by changing the variable in both Eqs. (1) and (3) using the definitions of Eqs. (9) and (10). After using some algebra, the final equations are for the integrand function:

$$F(U, T) = (2j+1) U \sigma_{n \to n'}(U) \left[ 1 - \left( \frac{\Delta E}{4U} \right)^2 \right] e^{-\frac{U}{k_B T} \left[ 1 + \left( \frac{\Delta E}{4U} \right)^2 \right]} \quad (17)$$

$$= (2j'+1) U \sigma_{n' \to n}(U) \left[ 1 - \left( \frac{\Delta E}{4U} \right)^2 \right] e^{-\frac{U}{k_B T} \left[ 1 + \left( \frac{\Delta E}{4U} \right)^2 \right]}$$

and for the rate coefficients:

$$k_{n \to n'}(T) = \frac{v_{ave}(T)}{(k_B T)^2} \left( \frac{e^{+\frac{\Delta E}{2k_B T}}}{2j+1} \right) \int_{U=\frac{\Delta E}{4}}^{\infty} F(U, T) dU, \quad (18)$$

$$k_{n' \to n}(T) = \frac{v_{ave}(T)}{(k_B T)^2} \left( \frac{e^{-\frac{\Delta E}{2k_B T}}}{2j'+1} \right) \int_{U=\frac{\Delta E}{4}}^{\infty} F(U, T) dU. \quad (19)$$

Comparing these formulae to Eqs. (12-13) and Eqs. (15-16) above one notices that this version is more symmetric because each formula has its own pre-factor. Also note that the range of integration is from $U = \Delta E/4$, which is the threshold for $U$, in agreement with the definition of Billing. Also, many factors in the two lines of in Eq. (17) are identical, except cross sections for excitation and quenching and their corresponding degeneracies. Therefore, in this symmetrized approach the reversibility is satisfied if $(2j+1)\sigma_{n \to n'}(U) = (2j'+1)\sigma_{n' \to n}(U)$, which is the simplest form of the principle. In Fig. 1 we plotted the values of $(2j'+1)\sigma_{n' \to n}$ vs $(2j+1)\sigma_{n \to n'}$ from MQCT calculations of Ref. (Mandal and Babikov, 2023) for $H_2O + H_2O$ at two values of energy: $U = 133$ and $708$ cm$^{-1}$. We see that at higher collision energies the reversibility is approximately satisfied, while at lower collision energies the deviations are more substantial.

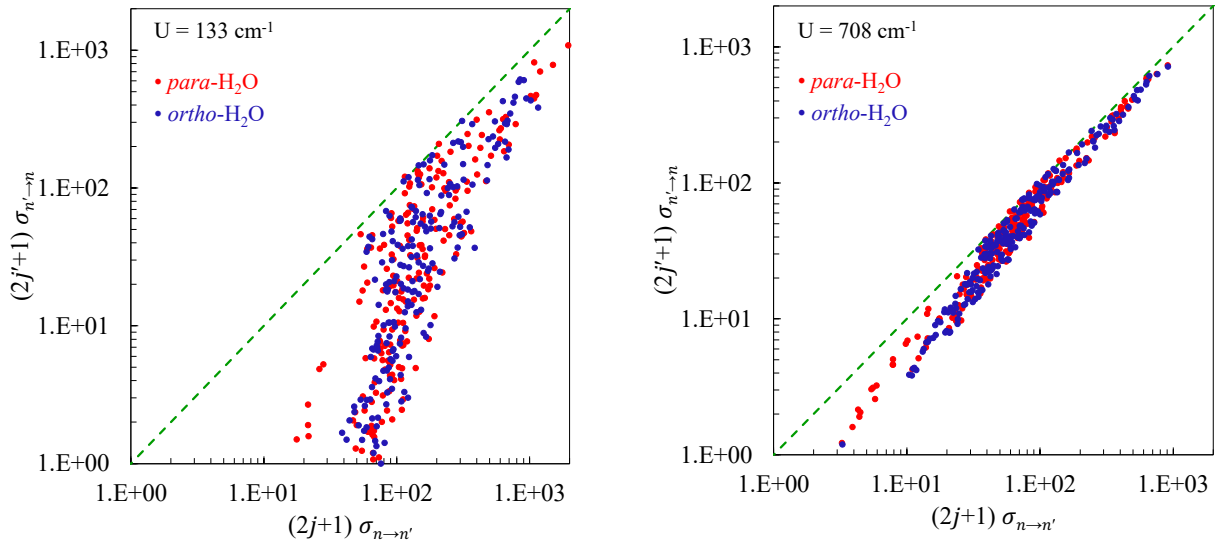

**Figure 1**: Comparison of $H_2O + H_2O$ collision cross sections (the units are Å$^2$) for the excitation and quenching directions of 441 state-to-state transitions at two values of collision energy $U$. The departure of datapoints from the diagonal line (dashed) signifies the deviation from the principle of microscopic reversibility.

## 2d. A unitless approach (Method 4)

In practice, when the fitting of the data is needed, Method 1 may have an advantage over Method 2, because the range of $F(E,T)$ is the same $E > 0$ for all transitions $n \leftrightarrow n'$, while the range of $F(E',T)$ determined by $E' > \Delta E$ is different for different transitions, simply because different transitions have different values of $\Delta E$. Method 3 also has different ranges for different transitions, $U > \Delta E/4$, but this can be easily avoided as described below.

Let's introduce a unitless variable $u = 4U/\Delta E$, for each transition independently. In a sense, $u$ measures the value of $U$ in the units of its minimal value, $\Delta E/4$. Then, the range of this unitless variable is always the same, $u > 1$. Also, lets introduce another unitless variable that measures $k_B T$ in the units of $\Delta E/4$, namely: $\varepsilon = 4k_B T/\Delta E$. Using these definitions, Eqs. (17-19) can be rewritten as follows:

$$F(u,\varepsilon) = (2j+1)\sigma_{n\to n'}(u)\left[u - \frac{1}{u}\right]e^{-\left[u+\frac{1}{u}\right]/\varepsilon} \qquad (20)$$

$$= (2j'+1)\sigma_{n'\to n}(u)\left[u - \frac{1}{u}\right]e^{-\left[u+\frac{1}{u}\right]/\varepsilon}$$

and for the rate coefficients:

$$k_{n\to n'}(T) = \frac{v_{ave}(T)}{\varepsilon^2}\left(\frac{e^{+2/\varepsilon}}{2j+1}\right)\int_{u=1}^{\infty} F(u,\varepsilon)du, \qquad (21)$$

$$k_{n'\to n}(T) = \frac{v_{ave}(T)}{\varepsilon^2}\left(\frac{e^{-2/\varepsilon}}{2j'+1}\right)\int_{u=1}^{\infty} F(u,\varepsilon)du. \qquad (22)$$

In these expressions all variables are unitless, except cross sections and collision velocities. Importantly, the integration range is independent of $\Delta E$. It is easy to see that in the high energy limit, $u \to \infty$, the expression for integrand simplifies and looks similar to that of Eqs. (1) and (3), e.g.:

$$F(u,\varepsilon) \sim (2j+1)u\sigma_{n\to n'}(u)e^{-u/\varepsilon}$$

$$\sim (2j'+1)u\sigma_{n'\to n}(u)e^{-u/\varepsilon}.$$

This unitless formulation may be advantageous for some applications.

## 3. Results and discussion

In what follows, we use Method 3 to compute temperature dependent rate coefficients for excitation and quenching. From the previous work (Mandal and Babikov, 2023) the values of cross sections are available at six collision energies: $U = 133, 200, 267, 400, 533$ and $708$ cm$^{-1}$. These are used to compute six values of the integrand function $F(U,T)$, at each temperature. The overall range of integration in Eqs. (18) and (19) is split onto three intervals treated differently:

$$U_0 < U < 200 \text{ cm}^{-1},$$

$$200 \leq U \leq 533 \text{ cm}^{-1},$$

$$533 < U < +\infty.$$

Here $U_0 = \Delta E/4$ is a threshold where $F(U,T) = 0$. The values of threshold are different for different transitions $n \leftrightarrow n'$ and in the database they range from $U_0 = 0.05$ cm$^{-1}$ for transition $5_{23} \leftrightarrow 6_{16}$ to $U_0 = 166$ cm$^{-1}$ for transition $6_{33} \leftrightarrow 0_{00}$.

In the low-energy part of the range (from threshold up to the second datapoint) the integral is computed analytically, by constructing a three-point analytic interpolation of the first three datapoints using the following function:

$$F_0(U) = \left(\frac{U-U_0}{C}\right)^A e^{-\frac{U-U_0}{B}}. \tag{23}$$

Expressions for the fitting coefficients $A$, $B$ and $C$ are given in Appendix A. In the middle part of the range, the integration is carried out numerically by constructing a cubic spline of the datapoints and using a constant-step quadrature with large number of points. In the high-energy interval (from the fifth datapoint to infinity) the integral is also computed analytically, by constructing the analytic fit of the last three datapoints using the following function:

$$F_\infty(U) = a\, e^{-\frac{U}{b} - \left(\frac{U}{c}\right)^2}, \tag{24}$$

and using this function to extrapolate into the range of large values of $U$. Expressions for the fitting coefficients $a$, $b$ and $c$ are also given in Appendix A.

From Eqs. (17-19) it follows that, having in our disposal cross sections $\sigma_{n \to n'}(U)$ for quenching and $\sigma_{n' \to n}(U)$ for excitation, we can, in principle, calculate two sets of rate coefficients. Namely, using the first line of Eq. (17) to compute $F(U)$ from $\sigma_{n \to n'}(U)$, and then using Eqs. (18-19) to compute $k_{n \to n'}(T)$ and $k_{n' \to n}(T)$, one can obtain the full set of rate coefficients for quenching and excitation from quenching cross sections alone. Alternatively, using the second line of Eq. (17) to compute $F(U)$ from $\sigma_{n' \to n}(U)$, and then using Eqs. (18-19) to compute $k_{n \to n'}(T)$ and $k_{n' \to n}(T)$, one can obtain another full set of rate coefficients for quenching and excitation from excitation cross sections alone. If the microscopic reversibility is satisfied, the two sets of data should be identical.

We did such calculations and compared two nascent sets of rate coefficients in Fig. 2, for $T = 300$ K. One can see that the rate coefficients obtained from the excitation and quenching cross sections are overall similar but not exactly the same, which means that the principle of microscopic reversibility is not exactly satisfied.

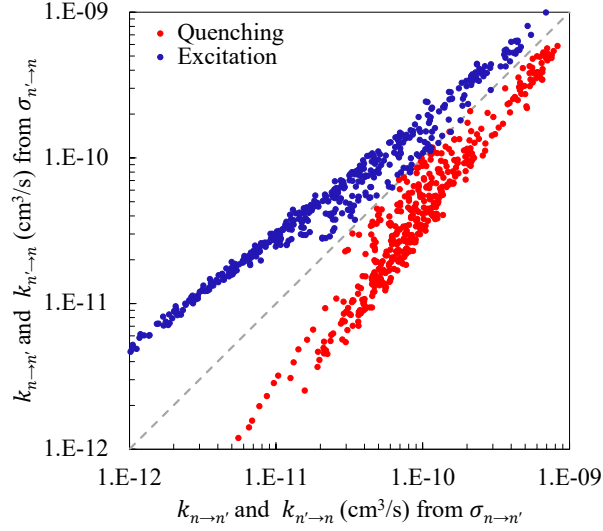

**Figure 2**: Comparison of two sets of rate coefficients obtained from quenching and excitation cross sections, as explained in the text. Departure of the datapoints from the diagonal line signifies deviation from the principle of microscopic reversibility. Red and blue color correspond to quenching and excitation rate coefficients, respectively.

In general, there is no strong argument in favor of one of these two sets of rate coefficients. However, one can notice that rate coefficients for excitation and quenching form similar patterns on two sides of the diagonal line in Fig. 2, which suggests that the desired (actual) rate coefficients must be somewhere between the two sets. Therefore, it makes sense to use both excitation and quenching cross sections at the same time to produce one set of rate coefficients. This can be done by constructing, for each transition $n \leftrightarrow n'$ at each temperature, the integrand function $\tilde{F}(U,T)$ that represents an average of the first and second lines of Eq (17), namely:

$$\tilde{F}(U,T) = \frac{(2j+1)\sigma_{n\to n'}(U) + (2j'+1)\sigma_{n'\to n}(U)}{2} \times U\left[1 - \left(\frac{\Delta E}{4U}\right)^2\right] e^{-\frac{U}{k_B T}\left[1+\left(\frac{\Delta E}{4U}\right)^2\right]} \quad (25)$$

Note that this formula includes both quenching and excitation cross sections $\sigma_{n\to n'}(U)$ and $\sigma_{n'\to n}(U)$ taken as a weighted sum to generate one single integrand function. The microscopic reversibility is automatically satisfied by Eqs. (18) and (19). Figure 3 gives examples of the integrand functions $\tilde{F}(U)$ constructed for three different values of temperature for all 441 transitions (ortho and para combined) in a broad energy range (from thresholds $U_0$ to $U \sim 10^4$ cm$^{-1}$). One can see that these functions grow quickly at threshold, pass through the computed datapoints smoothly, and then decay exponentially at high energy. This behavior is typical for many other molecule + atom and molecule + molecule systems.

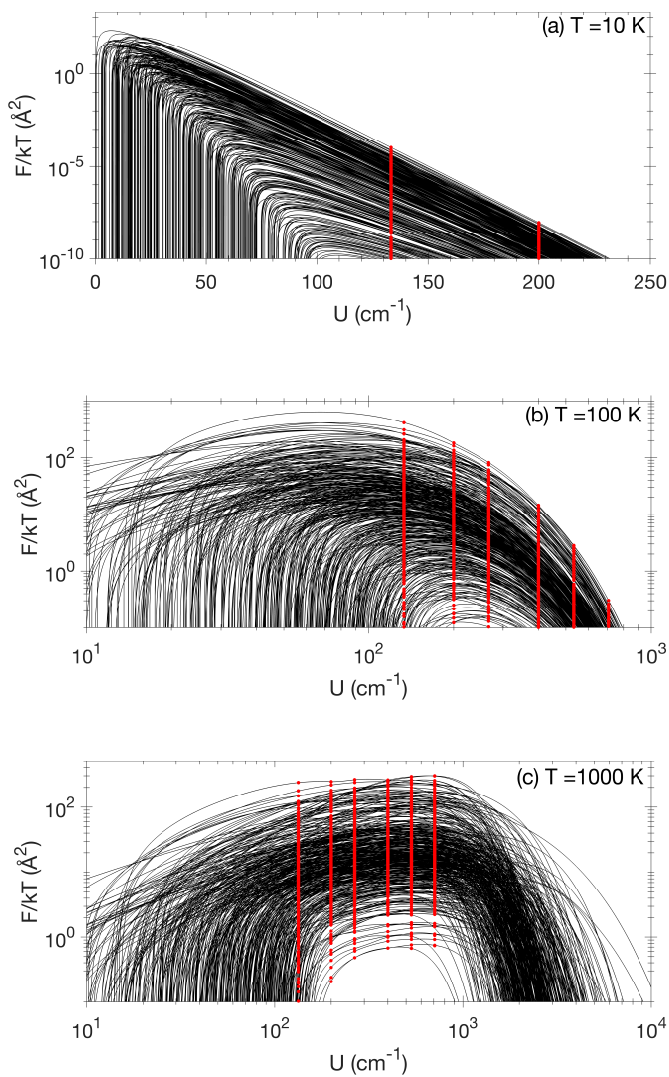

**Figure 3**: Energy dependence of average integrand functions $\tilde{F}(U)$ from Eq. (23) for all state-to-state transitions in the H$_2$O + H$_2$O database, at three values of temperature. Red symbols indicate MQCT datapoints at six values of collision energy $U$. Solid lines are used for analytic fits and numeric splines trough the temperature range.

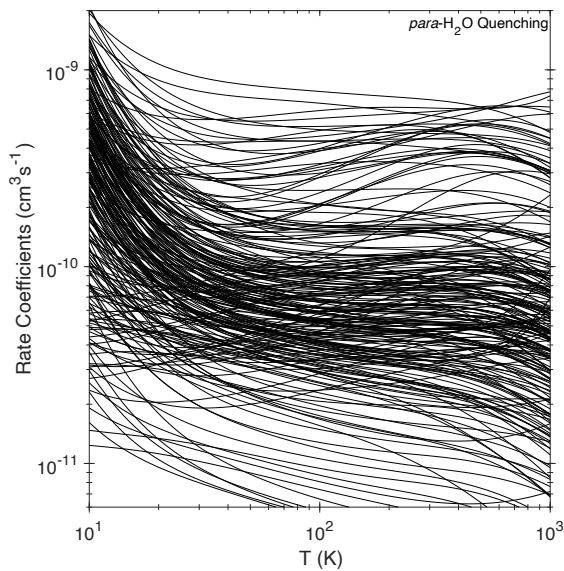
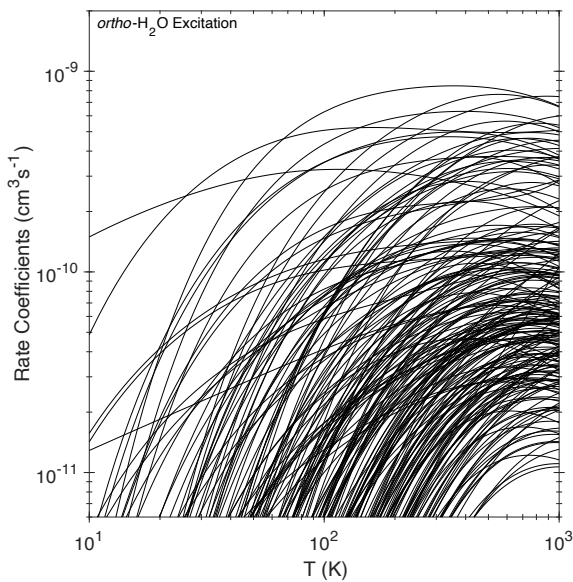
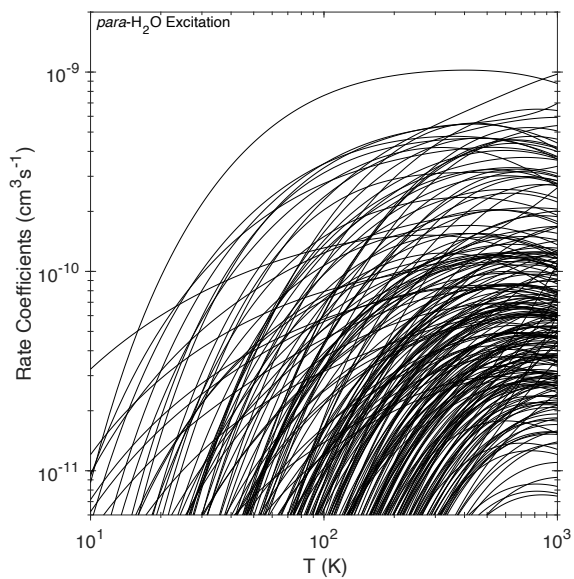
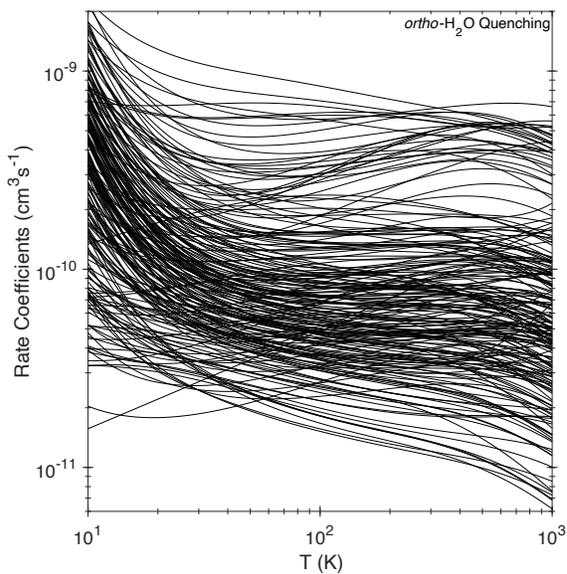

**Figure 4**: Temperature dependence of rate coefficients for 231 individual state-to-state transitions in the "target" *para*-$H_2O$ molecule. For the "collider" $H_2O$ molecule a thermal distribution of both *para*- and *ortho*-states is assumed.

**Figure 5**: Temperature dependence of rate coefficients for 210 individual state-to-state transitions in a "target" *ortho*-$H_2O$ molecule. For the "collider" $H_2O$ molecule a thermal distribution of both *para*- and *ortho*-states is assumed.

Figures 4 and 5 report the temperature dependencies of rate coefficients for individual transitions in $H_2O + H_2O$ obtained by integrating $\tilde{F}(U)$ as explained in Appendix A, and then using Eqs. (18) and (19) to ensure microscopic reversibility. Two frames of Fig. 4 report excitation and quenching rate coefficients for 231 transitions between para-states of water, while two frames of Fig. 5 report 210 transitions between ortho-stater of water, all in the range from 10 K to 1000 K. One can see that temperature dependencies of excitation cross sections are stronger, with many transitions characterized by cross sections that decrease exponentially when the temperature is reduced. In contrast, the rate coefficients for quenching usually increase at low temperature and vary less through the covered range.

In Figs. 6 and 7 we present a comparison of the rate coefficients computed here vs our previous work (Mandal and Babikov, 2023) and vs those from the work of other authors (Boursier et al., 2020; Buffa et al., 2000). Since here we re-use MQCT cross sections from our previous work, the differences between our new and old results are not expected to be large. Indeed, in Figs. 6 and 7 we see that red dots do not deviate significantly from the diagonal line. Still, we see that the deviations become more noticeable at high temperatures. For example, in *para*-$H_2O$ at $T$ = 200 K an RMS deviation of our new rate coefficients from old once is ~ 12% for stronger dipole-driven transitions and is ~ 18% for weaker transitions. At $T$ = 800 K these differences increase to ~ 30% for stronger and ~ 32% for weaker transitions. In *ortho*-$H_2O$ at $T$ = 200 K the RMS difference is ~ 11% for stronger transitions and is ~ 16% for weaker transitions. At $T$ = 800 K they increase to ~ 28% and ~ 29%, respectively. Therefore, our new approach, which includes integration over the range of collision energies, is recommended, in particular for higher temperatures.

It was explained in detail in our previous paper that for the stronger dipole-driven transitions our MQCT cross sections are somewhat smaller than those obtained by semi-classical methods of either (Buffa et al., 2000) or (Boursier et al., 2020), while for the weaker transitions our MQCT cross sections are somewhat larger than those of (Boursier et al., 2020). This is also true for a set of new rate coefficients obtained here. Indeed, in the range of larger rate coefficients of Figs. 6 and 7 (over $2 \times 10^{-10}$ cm$^3$/s) blue and cyan symbols concentrate *above* the diagonal line, while in the range of smaller rate coefficients (under $2 \times 10^{-10}$ cm$^3$/s) blue symbols are found *below* the diagonal line. Moreover, our new rate coefficients are somewhat smaller than our old rate coefficients, as characterized by RMS deviations quoted in the previous paragraph. Therefore, the difference between our results and those of either (Buffa et al., 2000) or (Boursier et al., 2020), is slightly higher here, compared to our previous work (Mandal and Babikov, 2023), and more so at higher temperatures.

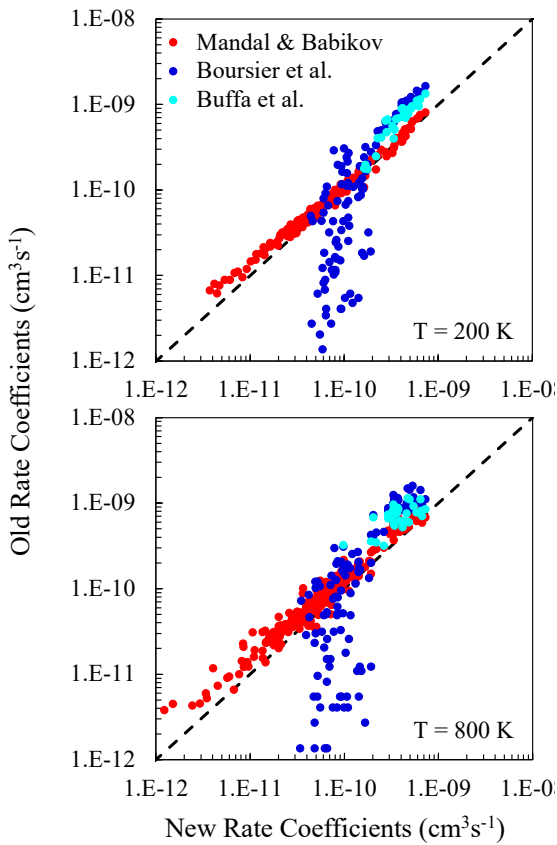 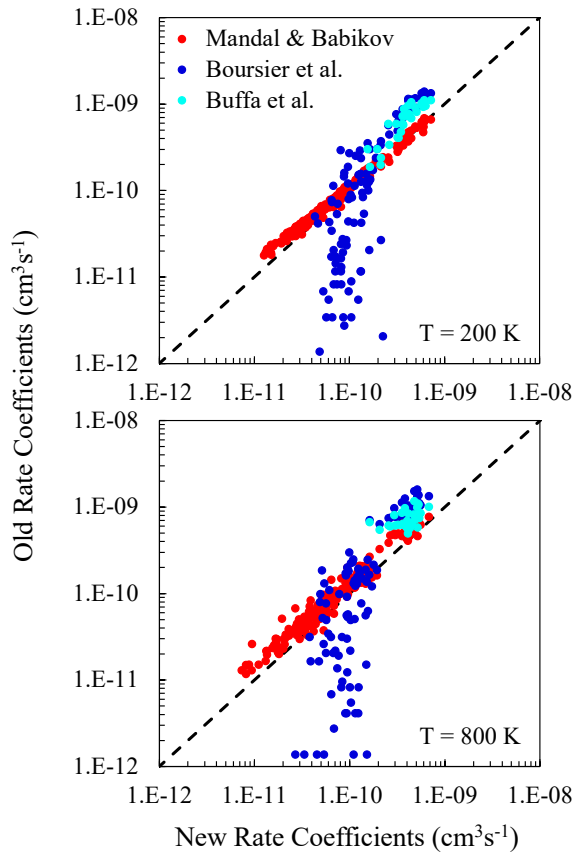

**Figure 6**: Comparison of rate coefficients computed here *vs* those available from literature for the quenching of *para*-$H_2O$ states by collisions with a thermal ensemble of $H_2O$ molecules at two values of temperature.

**Figure 7**: Comparison of rate coefficients computed here *vs* those available from literature for the quenching of *ortho*-$H_2O$ states by collisions with a thermal ensemble of $H_2O$ molecules at two values of temperature.

## 4. Conclusions

In this paper we described a new procedure for computing rate coefficients for the collision induced transitions between the states of molecules (rotational, vibrational) in a situation when the state-to-state transition cross sections are computed by an approximate method that does not satisfy the principle of microscopic reversibility automatically. This approach is based on the method of Billing, where an effective collision energy $U$ is introduced for the calculation of both excitation and quenching directions of the same transition (instead of dealing with two different collision energies for the excitation and quenching processes). The final formula of the symmetrized approach proposed here, Eqs. (17-19), are new to our best knowledge. They permit to use a weighted average of excitation and quenching cross sections for the prediction of rate coefficients that satisfy the principle of microscopic reversibility. In addition, they help to interpolate cross sections between the computed datapoints and the process threshold, which permits to expand the prediction of rate coefficient into the low temperature regime.

This method was applied to revise the existing database of rate coefficients for the rotational state-to-state transitions in $H_2O + H_2O$ collisions. In the older databases for $H_2O + H_2O$ the integration over collision energy was not done at all. Instead, the rate coefficient was predicted approximately using one value of cross section (at the collision energy that corresponds to the average collision speed at given temperature). Here, the actual integration over collision energy was carried out from the process threshold to infinity, using extrapolation of computed cross sections into the high-energy regime. As result, the range of temperatures was expanded from $150 \leq T \leq 800$ K in the older database, to $T \leq 1000$ K in the revised database. The data presented in the paper were plotted for the temperatures above 10 K, but we also looked at the predictions of rate coefficients down to $T \sim 5$ K and they seem to be reasonable and follow the same trend. These data can be used for the modeling of cometary comas and planetary atmospheres where $H_2O + H_2O$ collisions are important. For the potential users of this database, we created a computer code that generates rate coefficients for 231 transitions between para-states of water and 210 transitions between ortho-stater of water (in both excitation and quenching directions) at any given temperature in the range $5 \leq T \leq 1000$ K.

Of course, the low temperature predictions are based on the asymptotic form of cross-section behavior that has no information about such low-temperature quantum phenomena as scattering resonances, that may be important for $H_2O + H_2O$. For these to be described accurately one would have to do the full-quantum calculations of inelastic scattering, which is computationally unaffordable. The data available for $H_2O + H_2O$ system at this time come either from a semi-classical method (Boursier et al., 2020; Buffa et al., 2000), or from the mixed quantum/classical theory (Mandal et al., 2023; Mandal and Babikov, 2023), that are both approximate.

It should also be noted that all databases of rate coefficients for $H_2O + H_2O$ system developed so far involve averaging over rotational states of "collider" $H_2O$ molecules, assuming Boltzman distribution at given temperature. This is done, partially, due to tradition, and partially because the number of states in water is very large. For example, the present database reports 882 transitions in the "target" molecule if both excitation and quenching processes of para- and ortho-water are combined. If one chose not to average over the states of "collider" molecules, then each transition in the "target" needs to be combined with each transition in the "collider", leading to over 3/4 million (!) individual transitions in $H_2O + H_2O$ system. Even if such cross-section data are available, the energy dependence of cross sections for each transition must be fitted and integrated in an automatic but controllable manner, to avoid mistakes. And then, somehow be reported. The traditional way of reporting rate coefficients as an ASCI file may not be appropriate anymore. Thus, in the future, we would like to offer a database of the individual state-to-state transitions in the $H_2O + H_2O$ system resolved in the initial and final states of both collision partners, and available as a convenient subroutine that generates rate coefficients. This would give the indispensable input data, not available at present time, for the radiation transfer modelling in the $H_2O$ reach environments at conditions that are far from the local thermodynamic equilibrium, which is rather typical in astrophysics.

## Acknowledgements

This research was supported by NASA Astrophysics Program, grant number NNX17AH16G. DB acknowledges the support of Way Klingler Research Fellowship. BM acknowledges the support of Denis J. O'Brien and Eisch Fellowships. We used resources of the National Energy Research Scientific Computing Center, which is supported by the Office of Science of the U.S. Department of Energy under Contract No. DE-AC02-5CH11231. This research also used HPC resources at Marquette funded in part by the National Science foundation award CNS-1828649. Marie-Lise Dubernet at the Observatory of Paris is gratefully acknowledged for her continuous support of these efforts. Martin Cordiner at NASA is acknowledged for stimulating discussions and his encouragement of this project.

**References**

Al-Edhari, A.J., Ceccarelli, C., Kahane, C., Viti, S., Balucani, N., Caux, E., Faure, A., Lefloch, B., Lique, F., Mendoza, E., Quenard, D., Wiesenfeld, L., 2017. Astron Astrophys 597, A40.

Billing, G.D., 1984. Computer Physics Reports 1, 239–296.

Billing, G.D., 2003. The quantum classical theory. Oxford University Press.

Bockelée-Morvan, D., Hartogh, P., Crovisier, J., Vandenbussche, B., Swinyard, B.M., Biver, N., Lis, D.C., Jarchow, C., Moreno, R., Hutsemékers, D., Others, 2010. Astron Astrophys 518, L149.

Bonev, B.P., dello Russo, N., DiSanti, M.A., Martin, E.C., Doppmann, G., Vervack, R.J., Villanueva, G.L., Kawakita, H., Gibb, E.L., Combi, M.R., Roth, N.X., Saki, M., McKay, A.J., Cordiner, M.A., Bodewits, D., Crovisier, J., Biver, N., Cochran, A.L., Shou, Y., Khan, Y., Venkataramani, K., 2021. Planetary Science Journal 2, 45.

Boursier, C., Mandal, B., Babikov, D., Dubernet, M.L., 2020. Mon Not R Astron Soc 498, 5489–5497.

Bruzzone, L., Plaut, J.J., Alberti, G., Blankenship, D.D., Bovolo, F., Campbell, B.A., Ferro, A., Gim, Y., Kofman, W., Komatsu, G., McKinnon, W., Mitri, G., Orosei, R., Patterson, G.W., Plettemeier, D., Seu, R., 2013. RIME: Radar for Icy moon Exploration, in: International Geoscience and Remote Sensing Symposium (IGARSS).

Buffa, G., Tarrini, O., Scappini, F., Cecchi-Pestellini, C., 2000. Astrophys J Suppl Ser 128, 597–601.

Cochran, A.L., Levasseur-Regourd, A.-C., Cordiner, M., Hadamcik, E., Lasue, J., Gicquel, A., Schleicher, D.G., Charnley, S.B., Mumma, M.J., Paganini, L., Others, 2015. Space Sci Rev 197, 9–46.

Cordiner, M.A., Coulson, I.M., Garcia-Berrios, E., Qi, C., Lique, F., Zołtowski, M., de Val-Borro, M., Kuan, Y.-J., Ip, W.-H., Mairs, S., Roth, N.X., Charnley, S.B., Milam, S.N., Tseng, W.-L., Chuang, Y.-L., 2022. Astrophys J 929, 38.

Daniel, F., Dubernet, M.-L., Grosjean, A., 2011. Astron Astrophys 536, A76.

Dones, L., Brasser, R., Kaib, N., Rickman, H., 2015. Space Sci Rev 197, 191–269.

Dubernet, M.L., Quintas-Sánchez, E., 2019. Mol Astrophys 16.

Enya, K., Kobayashi, M., Kimura, J., Araki, H., Namiki, N., Noda, H., Kashima, S., Oshigami, S., Ishibashi, K., Yamawaki, T., Tohara, K., Saito, Y., Ozaki, M., Mizuno, T., Kamata, S., her continuous support of these efforts. Martin Cordiner at NASA is acknowledged for stimulating discussions and his encouragement of this project.

**References**

Al-Edhari, A.J., Ceccarelli, C., Kahane, C., Viti, S., Balucani, N., Caux, E., Faure, A., Lefloch, B., Lique, F., Mendoza, E., Quenard, D., Wiesenfeld, L., 2017. Astron Astrophys 597, A40.

Billing, G.D., 1984. Computer Physics Reports 1, 239–296.

Billing, G.D., 2003. The quantum classical theory. Oxford University Press.

Bockelée-Morvan, D., Hartogh, P., Crovisier, J., Vandenbussche, B., Swinyard, B.M., Biver, N., Lis, D.C., Jarchow, C., Moreno, R., Hutsemékers, D., Others, 2010. Astron Astrophys 518, L149.

Bonev, B.P., dello Russo, N., DiSanti, M.A., Martin, E.C., Doppmann, G., Vervack, R.J., Villanueva, G.L., Kawakita, H., Gibb, E.L., Combi, M.R., Roth, N.X., Saki, M., McKay, A.J., Cordiner, M.A., Bodewits, D., Crovisier, J., Biver, N., Cochran, A.L., Shou, Y., Khan, Y., Venkataramani, K., 2021. Planetary Science Journal 2, 45.

Boursier, C., Mandal, B., Babikov, D., Dubernet, M.L., 2020. Mon Not R Astron Soc 498, 5489–5497.

Bruzzone, L., Plaut, J.J., Alberti, G., Blankenship, D.D., Bovolo, F., Campbell, B.A., Ferro, A., Gim, Y., Kofman, W., Komatsu, G., McKinnon, W., Mitri, G., Orosei, R., Patterson, G.W., Plettemeier, D., Seu, R., 2013. RIME: Radar for Icy moon Exploration, in: International Geoscience and Remote Sensing Symposium (IGARSS).

Buffa, G., Tarrini, O., Scappini, F., Cecchi-Pestellini, C., 2000. Astrophys J Suppl Ser 128, 597–601.

Cochran, A.L., Levasseur-Regourd, A.-C., Cordiner, M., Hadamcik, E., Lasue, J., Gicquel, A., Schleicher, D.G., Charnley, S.B., Mumma, M.J., Paganini, L., Others, 2015. Space Sci Rev 197, 9–46.

Cordiner, M.A., Coulson, I.M., Garcia-Berrios, E., Qi, C., Lique, F., Zołtowski, M., de Val-Borro, M., Kuan, Y.-J., Ip, W.-H., Mairs, S., Roth, N.X., Charnley, S.B., Milam, S.N., Tseng, W.-L., Chuang, Y.-L., 2022. Astrophys J 929, 38.

Daniel, F., Dubernet, M.-L., Grosjean, A., 2011. Astron Astrophys 536, A76.

Dones, L., Brasser, R., Kaib, N., Rickman, H., 2015. Space Sci Rev 197, 191–269.

Dubernet, M.L., Quintas-Sánchez, E., 2019. Mol Astrophys 16.

Enya, K., Kobayashi, M., Kimura, J., Araki, H., Namiki, N., Noda, H., Kashima, S., Oshigami, S., Ishibashi, K., Yamawaki, T., Tohara, K., Saito, Y., Ozaki, M., Mizuno, T., Kamata, S.,

Matsumoto, K., Sasaki, S., Kuramoto, K., Sato, Y., Yokozawa, T., Numata, T., Mizumoto, S., Mizuno, H., Nagamine, K., Sawamura, A., Tanimoto, K., Imai, H., Nakagawa, H., Kirino, O., Green, D., Fujii, M., Iwamura, S., Fujishiro, N., Matsumoto, Y., Lingenauber, K., Kallenbach, R., Althaus, C., Behnke, T., Binger, J., Daurskikh, A., Eisenmenger, H., Heer, U., Hüttig, C., Lara, L.M., Lichopoj, A., Lötzke, H.G., Lüdicke, F., Michaelis, H., Pablo Rodriguez Garcia, J., Rösner, K., Stark, A., Steinbrügge, G., Thabaut, P., Thomas, N., del Togno, S., Wahl, D., Wendler, B., Wickhusen, K., Willner, K., Hussmann, H., 2022. Advances in Space Research 69, 2283–2304.

Faure, A., Josselin, E., 2008. Astron Astrophys 492, 257–264.

Faure, A., Lique, F., Loreau, J., 2020. Mon Not R Astron Soc 493, 776–782.

Faure, A., Lique, F., Remijan, A.J., 2018. Journal of Physical Chemistry Letters 9, 3199–3204.

Faure, A., Wiesenfeld, L., Wernli, M., Valiron, P., 2005. J Chem Phys 123, 104309.

Grasset, O., Dougherty, M.K., Coustenis, A., Bunce, E.J., Erd, C., Titov, D., Blanc, M., Coates, A., Drossart, P., Fletcher, L.N., Hussmann, H., Jaumann, R., Krupp, N., Lebreton, J.P., Prieto-Ballesteros, O., Tortora, P., Tosi, F., van Hoolst, T., 2013. JUpiter ICy moons Explorer (JUICE): An ESA mission to orbit Ganymede and to characterise the Jupiter system. Planet Space Sci.

Hartogh, P., Lellouch, E., Moreno, R., Bockelée-Morvan, D., Biver, N., Cassidy, T., Rengel, M., Jarchow, C., Cavalié, T., Crovisier, J., Others, 2011. Astron Astrophys 532, L2.

Joy, C., Mandal, B., Bostan, D., Babikov, D., 2023. PCCP submitted.

ben Khalifa, M., Quintas-Sánchez, E., Dawes, R., Hammami, K., Wiesenfeld, L., 2020. Phys Chem Chem Phys 22, 17494–17502.

Loreau, J., Faure, A., Lique, F., 2018. J Chem Phys 148, 244308.

Mandal, B., Babikov, D., 2023. Astron Astrophys 671, A51.

Mandal, B., Joy, C., Bostan, D., Eng, A., Babikov, D., 2023. Journal of Physical Chemistry Letters 14, 817–824.

Mandal, B., Joy, C., Semenov, A., Babikov, D., 2022. ACS Earth Space Chem 6, 521–529.

Mandal, B., Semenov, A., Babikov, D., 2020. Journal of Physical Chemistry A 124, 9877–9888.

Roth, N.X., Bonev, B.P., DiSanti, M.A., Russo, N. dello, McKay, A.J., Gibb, E.L., Saki, M., Khan, Y., Vervack, R.J., Kawakita, H., Cochran, A.L., Biver, N., Cordiner, M.A., Crovisier, J., Jehin, E., Weaver, H., 2021. Planetary Science Journal 2, 54.

Semenov, A., Babikov, D., 2013. J Phys Chem Lett 5, 275–278.

Semenov, A., Babikov, D., 2014. J Chem Phys 140, 44306.

Semenov, A., Babikov, D., 2015a. Journal of Physical Chemistry A 119.

Semenov, A., Babikov, D., 2015b. J Phys Chem Lett 6, 1854–1858.

Semenov, A., Babikov, D., 2016. J Phys Chem A 120, 3861–3866.

Semenov, A., Dubernet, M.L., Babikov, D., 2014. Journal of Chemical Physics 141, 114304.

Semenov, A., Ivanov, M., Babikov, D., 2013. J Chem Phys 139, 74306.

Sun, Z.-F., van Hemert, M.C., Loreau, J., van der Avoird, A., Suits, A.G., Parker, D.H., 2020. Science (1979) 369, 307–309.

Vorburger, A., Fatemi, S., Galli, A., Liuzzo, L., Poppe, A.R., Wurz, P., 2022. Icarus 375, 114810.

Wiesenfeld, L., 2021. J Chem Phys 155, 71104.

Wiesenfeld, L., Faure, A., 2010. Phys Rev A (Coll Park) 82, 40702.

Wiesenfeld, L., Scribano, Y., Faure, A., 2011. Physical Chemistry Chemical Physics 13, 8230–8235.

Wirström, E.S., Bjerkeli, P., Rezac, L., Brinch, C., Hartogh, P., 2020. Astron Astrophys 637, A90.

Żółtowski, M., Loreau, J., Lique, F., 2022. Physical Chemistry Chemical Physics 24, 11910–11918.

**Appendix A: The fitting formulae**

*Low energy range:* Using three data points with the values of energy $U_1$, $U_2$ and $U_3$ and the corresponding values of the function $F_1$, $F_2$ and $F_3$ at the low-energy side, the fitting coefficients for Eq. (23) are obtained as follows:

$$A = \frac{(U_3 - U_1)\ln(F_2/F_1) - (U_2 - U_1)\ln(F_3/F_1)}{(U_3 - U_1)\ln(U_2/U_1) - (U_2 - U_1)\ln(U_3/U_1)},$$

then

$$B = \frac{(U_2 - U_1)}{\ln\{(F_1/F_2)(U_2/U_1)^A\}},$$

and then

$$C = \frac{(U_1 - U_0)}{F_1^{1/A}} \exp\left\{\frac{U_1 - U_0}{AB}\right\}.$$

Unfortunately, analytic integration of $\tilde{F}_0(U)$ in the form of Eq. (23) requires calculations of the incomplete gamma-function $\gamma(A + 1, U_2/B)$, see Gradshteyn & Ryzhik (*Table of Integrals, Series and Products*). To avoid this complication, we set $A = 1$ which is equivalent to using a two-point fitting formula. Then, using coefficients $B$ and $C$ only, the integral of $\tilde{F}(U)$ in the range from $U_0$ to $U_2$ is computed easily:

$$\int_{U_0}^{U_2} \tilde{F}_0(U) dU = \int_{U_0}^{U_2} \left(\frac{U - U_0}{C}\right) e^{-\frac{U - U_0}{B}} dU = \frac{B^2}{C}\left[1 - \left(\frac{C}{U_2 - U_0} + \frac{C}{B}\right) F_2\right].$$

*High energy range:* Similar, using three data points with the values of energy $U_4$, $U_5$ and $U_6$ and the corresponding values of the function $F_4$, $F_5$ and $F_6$ at the high-energy side, the coefficients for Eq. (24) are obtained as follows:

$$b = \frac{(U_5 - U_4)(U_6 - U_5)(U_6 - U_4)}{(U_6^2 - U_4^2)\ln(F_6/F_5) - (U_6^2 - U_5^2)\ln(F_6/F_4)},$$

$$c = \sqrt{\frac{(U_5 - U_4)(U_6 - U_5)(U_6 - U_4)}{(U_6 - U_5)\ln(F_6/F_4) - (U_6 - U_4)\ln(F_6/F_5)}},$$

then

$$a = F_6 \exp\left\{\frac{U_6}{b} + \left(\frac{U_6}{c}\right)^2\right\}.$$

It is assumed that $c^2 \geq 0$. At high temperature (near $T = 800$ K) for some transitions the last data point (at $U_6 = 708$ cm$^{-1}$) seems to be inaccurate (most likely due to thermal averaging over the states of quencher) which causes $c^2$ to be negative. In such cases, we disregard this last point and simply use the previous one. If $c^2$ is still negative (very rare), we set $c = +\infty$ and use a two-point fitting formula.

Using coefficients $a$, $b$ and $c$, the integral of $\tilde{F}(U)$ in the range from $U_5$ to $+\infty$ is computed analytically:

$$\int_{U_5}^{+\infty} \tilde{F}_\infty(U) dU = \int_{U_5}^{+\infty} a\, e^{-\frac{U}{b} - \left(\frac{U}{c}\right)^2} dU = \frac{\sqrt{\pi}}{2} ac \exp\left\{\frac{c^2}{4b^2}\right\}\left[1 - \mathrm{erf}\left\{\frac{2bU_5 + c^2}{2bc}\right\}\right]$$

The last two factors in this formula can turn very large and very small simultaneously, which causes practical difficulty for some transitions. A convenient way to deal with error function is to use an approximate expression:

$$\operatorname{erf} x = 1 - P\{x\}e^{-x^2}$$

where $P\{x\}$ is a fifth-order polynomial of the form given by Abramovitz & Stegun (*Handbook of Mathematical Functions with Formulas, Graphs and Mathematical Tables*). This permits to express the integral through the value of $F_5$ and the polynomial, as follows:

$$\int_{U_5}^{+\infty} \tilde{F}_\infty(U) dU = \frac{\sqrt{\pi}}{2} cF_5 P\left\{\frac{2bU_5 + c^2}{2bc}\right\}.$$

When the two-point extrapolation is used, the expression is:

$$\int_{U_5}^{+\infty} \tilde{F}_\infty(U) dU = \int_{U_5}^{+\infty} a\, e^{-\frac{U}{b}} dU = bF_5.$$